\documentclass[11pt]{article}
\usepackage[english]{babel}
\usepackage[latin1]{inputenc}	
\usepackage{amsmath}
\usepackage{amsfonts}
\usepackage{amssymb}
\usepackage[usenames,dvipsnames]{xcolor}

\usepackage{marginnote}

\usepackage{hyperref}

\DeclareMathOperator{\artanh}{artanh}

\newcommand{\mc}[1]{\mathcal{#1}}
\renewcommand{\d}{{\mathrm{d}}}
\newcommand{\w}{{\wedge}}
\newcommand{\n}{\nabla}

\newcommand{\La}{{\Lambda}}
\newcommand{\p}{{\partial}}
\newcommand{\al}{{\alpha}}
\newcommand{\be}{{\beta}}

\newcommand{\ep}{{\epsilon}}

\newcommand{\ga}{{\gamma}}

\newcommand{\la}{{\lambda}}
\newcommand{\si}{{\sigma}}

\newcommand{\om}{{\omega}}
\newcommand{\Om}{{\Omega}}

\newcommand{\lp}{\left(}
\newcommand{\rp}{\right)}
\newcommand{\lb}{\left[}
\newcommand{\rb}{\right]}
\newcommand{\BE}{\begin{equation}}
\newcommand{\EE}{\end{equation}}
\newcommand{\BSE}{\begin{subequations}}
\newcommand{\ESE}{\end{subequations}}
\newcommand{\BC}{\begin{center}}
\newcommand{\EC}{\end{center}}
\newcommand{\BF}{\begin{figure}}
\newcommand{\EF}{\end{figure}}

\begin{document}

\setlength{\unitlength}{1mm}
\thispagestyle{empty}


\BC
{\bf \LARGE Integrability in conformally coupled gravity:\\
\medskip
Taub-NUT spacetimes and rotating black holes}
\vspace{1.5cm}

{\bf Yannis Bardoux,}$^{1}\,$
{\bf Marco M.~Caldarelli}$^{2}\,$
{\bf and Christos Charmousis}$^{1,3}\,$

\vspace{0.5cm}

{\it $^1$\,Laboratoire de Physique Théorique (LPT), Univ. Paris-Sud,\\
 CNRS UMR 8627, F-91405 Orsay, France}

{\it $^2$\,Mathematical Sciences and STAG research centre, University of Southampton,\\
Highfield, Southampton SO17 1BJ, United Kingdom}

{\it $^3$\,Laboratoire de Mathématiques et Physique Théorique (LMPT), Univ. Tours,\\ UFR Sciences et Techniques,\\ Parc de Grandmont, F-37200 Tours, France}

\vspace{0.3cm} {\small\tt
yannis.bardoux@th.u-psud.fr,
m.m.caldarelli@soton.ac.uk,
christos.charmousis@th.u-psud.fr
}

\vspace{1cm} 
{\bf ABSTRACT}\\\EC
We consider four dimensional stationary and axially symmetric spacetimes for conformally coupled scalar-tensor theories. We show that, in analogy to the Lewis-Papapetrou problem in General Relativity (GR), the theory at hand can be recast in an analogous integrable form. We give the relevant rod formalism, introduced by Weyl for vacuum GR,  explicitly giving the rod structure of the black hole of Bocharova et al.\ and Bekenstein (BBMB), in complete analogy to the Schwarzschild solution. The additional scalar field is shown to play the role of an extra Weyl potential. We then employ the Ernst method as a concrete solution generating example to obtain the Taub-NUT version of the BBMB hairy black hole, with or without a cosmological constant. We show that the anti-de Sitter hyperbolic version of this solution is free of closed timelike curves that plague usual Taub-NUT metrics, and thus consists of a rotating, asymptotically locally anti-de Sitter black hole. This stationary solution has no curvature singularities whatsoever in the conformal frame, and the NUT charge is shown here to regularize the central curvature singularity of the corresponding static black hole. Given our findings we discuss the anti-de Sitter hyperbolic version of Taub-NUT in four dimensions, and show that the curvature singularity of the NUT-less solution is now replaced by a neighboring chronological singularity screened by horizons. We argue that the properties of this rotating black hole are very similar to those of the rotating BTZ black hole in three dimensions.

\newpage
\tableofcontents

\section{Introduction}

It is well known that solution generating techniques in General Relativity (GR) are very powerful, and seemingly difficult non-linear problems are often integrable (for a full discussion, see \cite{KSMH} and references therein). Typically one starts by considering spacetimes with special geometrical properties, such as Petrov type D metrics \cite{Plebanski:1976gy}, and as a result most common black hole spacetimes are found to belong to this class. Or again one assumes the presence of symmetries in the shape of Killing vectors, such as in the case of Weyl  static and axisymmetric spacetimes \cite{Weyl:1917gp}, finding nice analogies with ordinary Newtonian gravity (see \cite{Dowker:2001dg} for a concise overview), leading to numerous multi-black hole solutions. Assuming then stationary rather than static spacetimes, one obtains Lewis-Papapetrou metrics~\cite{Papapetrou:1953zz} where numerous formalisms, such as that of Ernst \cite{Ernst:1967wx}, enable the generation of rotating metrics from static ones (e.g.\ Kerr from Schwarzschild). 

Given a similar set of symmetries, an obvious question then is which of these solution-generating techniques can be generalized to higher dimensions and to more complex gravitating theories. This interest is manifest from recent advances in string gravity/holography and the need for a generic and concise way of obtaining solutions in the presence of a cosmological constant, with additional fields, and/or in higher dimensions. Furthermore, this interest is enhanced by four dimensional theories of gravity modification -- like scalar-tensor theories -- where exact solutions are in need in order to understand issues, such as no hair theorems, and to test them at strong gravity scales.  Many of the above mentioned integrability properties are however tied to four dimensional gravity. Typically, Petrov type D metrics allow the inclusion of a cosmological constant, as was shown by the pioneering work of Carter \cite{Carter:1968ks}, or an electromagnetic field, but one still does not know, for example, the equivalent of the charged rotating solution of the Kerr-Newman black hole in higher dimensions\footnote{In such cases one has to resort to analytical perturbative methods, such as the blackfold approach \cite{Emparan:2009cs}, that allowed the construction of approximate, higher-dimensional Kerr-Newman solutions, at least in some regimes \cite{Caldarelli:2010xz}.}. Furthermore, although some of the properties of the Weyl metrics survive in higher dimensions \cite{Emparan:2001wk}, it is found that the cosmological constant spoils the integrability properties of the Weyl and Papapetrou problems \cite{Charmousis:2003wm,Charmousis:2006fx} by introducing a non-trivial curvature scale in the action. In fact, it is found that the inclusion of a curvature scale does not permit a universal coordinate system in which integration techniques are possible. In other words, although generic methods, such as that of Ernst, go through \cite{Charmousis:2006fx}, one has to adapt them each time to the solution sought, making therefore novel solutions difficult to find. A notable exception to this negative statement has been given recently by Astorino \cite{Astorino:2012zm}. One can also  develop powerful analytical approximation techniques such as the matched asymptotic expansion \cite{Emparan:2007wm,Caldarelli:2008pz} and the aforementioned blackfold effective theory \cite{Emparan:2009cs} in order to study the geometric and physical properties of black holes whose exact analytic form is unknown.
In this article we will however restrict our attention to exact solutions of the field equations and to methods to generate them.

In practice, it turns out that the addition of massless matter fields enjoying some symmetry (possibly not shared by the full theory) sometimes provides better integrability properties than a simple cosmological constant, precisely because of the absence of an additional curvature scale. This is the case that we will study here, concerning four dimensional conformally coupled scalar-tensor theories.
To be more precise, let us consider the following gravitational action,
\begin{equation}
 S_0[g_{ab},\phi] =
 \int_\mc M \sqrt{-g} \lp \frac{R-2\La}{16\pi G} - \frac1{12}R\phi^2 - \frac12 \p_a \phi \p^a \phi - \al \phi^4 \rp \d^4 x  \label{action}
\end{equation}
where $G$ is the gravitational Newton constant, and we have included for later convenience the cosmological constant $\La$ and a self-interaction quartic potential with arbitrary coupling $\al$. It is also useful to introduce the effective gravitational constant
\BE \tilde{G}(\phi)=\lp 1-\frac{4\pi G}{3} \phi^2\rp ^{-1} G . \label{Gtilde}\EE
The variation of the action \eqref{action} with respect to the metric $g_{ab}$ gives
\BE
G_{ab} + \La g_{ab} = 8\pi G T_{ab}^\phi  \label{EOMg},
\EE
with the scalar field energy-momentum tensor
\BE  T_{ab}^\phi=\p_a \phi  \p_b \phi - \frac12 g_{ab} \p_c \phi \p^c \phi + \frac16 \lp g_{ab}\Box - \n_a \n_b + G_{ab} \rp \phi^2- \al g_{ab} \phi^4 . \EE
Variation with respect to the scalar field gives,
\BE \Box\phi = \frac16 R \phi + 4\al \phi^3 . \label{EOMphi} \EE
This latter equation is invariant under the conformal transformation \cite{Bekenstein:1974sf}
\BE \lp g_{ab}, \phi \rp \mapsto \lp \Om^2 g_{ab}, \Om^{-1} \phi \rp \EE
where $\Om$ is a smooth and strictly positive function on $\mc M$. Moreover, given this conformal symmetry, the trace of the energy-momentum tensor vanishes, $T_a{}^a=0$. As a result, the equation of motion for the scalar field \eqref{EOMphi} becomes
\BE \Box\phi = \frac23 \La \phi + 4\al \phi^3 . \label{EOMphi2}\EE
Taking $\Lambda=0$ and $\alpha=0$ gives us a massless theory, for which we will examine the Lewis-Papapetrou problem. The more general case will then be analysed in a later section.

This massless scalar-tensor theory has interesting, non-trivial, and quite simple solutions, starting with the static BBMB solution \cite{BBMB} found by Bocharova et al.\ and later rediscovered by Bekenstein \cite{Bekenstein:1974sf}, and its  interesting extension with a cosmological constant -- the MTZ black hole -- constructed by Martinez et al.~\cite{Martinez:2002ru}. Since these pioneering works, extensions of these black holes have been found \cite{Martinez:2005di, Charmousis:2009cm, Bardoux:2012tr} (see also \cite{Astorino:2013xc,Astorino:2013sfa} for C-metric solutions), most impressively including the general Petrov type D metric found by Anabalon and Maeda \cite{Anabalon:2009qt}. Scalar-tensor theories are in general an interesting laboratory of gravity modification and this particular theory, although not particularly phenomenologically oriented, gives non-trivial examples and has important integrability properties due to the conformal symmetry of the scalar. So, how far can we go within this theory? In this paper we will study the Lewis-Papapetrou system associated to the above action \eqref{action} for $\Lambda=0$ and $\alpha=0$. We will see that it is essentially an integrable problem, as it is in the case of GR, as long as the scalar is massless and we do not have a cosmological constant. As a concrete example, we will construct explicitly the Taub-NUT metric of this theory, and we will study its properties. 
It should be possible to recover it as a singular limit  of the Petrov type D metrics found in \cite{Anabalon:2009qt}, but we will not explicitly discuss this limit here, for the solution will be found using the Ernst method. 
Using the insight gained from massless case, the solution is easily extended 
to the case of $\Lambda \neq 0$ and $\alpha\neq 0$. The presence of the cosmological constant regularizes the solution, giving a rotating black hole geometry. This is the first explicit example of a rotating black hole within this theory.

Taub-NUT spacetime has been described by Misner as the counterexample to almost everything in gravity \cite{Misner:1965zz}.
In its original form it is a vacuum solution to Einstein's equations. A concise description of its most important properties is given in \cite{Hawking:1973uf}. The solution was discovered by Taub in 1951 in its cosmological chart \cite{Taub:1950ez}, and extended by Newman, Tamburino and Unti in 1963 in its static region \cite{Newman:1963yy}. The metric is usually written in the following form,
\BE \d s^2 = - \frac{r^2-2mr-n^2}{r^2+n^2} \lp \d t + 2n\cos\theta\,\d\varphi \rp^2 + \frac{r^2+n^2}{r^2-2mr-n^2} \d r^2 + (r^2+n^2) \lp \d\theta^2 + \sin^2\theta\,\d\varphi^2 \rp \label{taub-nut} \EE
where $n$ is the NUT parameter. Due to the presence of the parameter $n$, the solution is not asymptotically flat but also has no curvature singularity at $r=0$. The solution has two Killing horizons given by the radial coordinates $r_\pm=m\pm\sqrt{m^2+n^2}$.  In the region where the coordinate $t$ is spacelike, it was noted in \cite{Taub:1950ez} that this solution describes a Big Bang at $r=r_-$ and a Big Crunch at $r=r_+$, and that the spacelike slices have the topology of $S^3$. The stationary, spacelike NUT regions contain unfortunately closed timelike curves (CTCs) as a consequence of the periodicity of $t$, which is imposed in order to avoid Misner strings at $\theta=0$, $\pi$. The presence of CTCs generically discards the Lorentzian signature version of these metrics, and in general only the Euclidean signature solutions are studied, which can give nut and bolt instantons. The NUT parameter $n$ is associated to the notion of gravitational magnetic mass. In fact, sections of constant $r$ describe the geometry of a principal $U(1)$ fibration over $S^2$ where the coordinate $t$ has the periodicity $8\pi n$ (see e.g.~\cite{Eguchi:1980jx}). The geometric cross-term $\mc A = 2n\cos\theta\,\d\varphi$ in \eqref{taub-nut} is analogous to the potential of the electromagnetic field generated by a magnetic monopole (the Dirac monopole) of charge proportional to $n$, up to a gauge transformation. Thus, there is a dictionary between this $U(1)$-bundle over $S^2$ and a magnetic monopole, introduced by Dirac in order to explain the quantization of electric charge \cite{Dirac:1931kp}. In fact the periodicity imposed on the coordinate $t$ is analogous to the Dirac quantization condition of the electric charge.
Moreover, we see that the potential $\mc A$ cannot be globally defined: two patches have to be used to cover $S^2$. These singularities of the potential $\mc A$ at $\theta=0$ or $\theta=\pi$ are called Misner string singularities in the context of the Taub-NUT solution, and are analogous to the Dirac string singularities when we consider the field produced by a magnetic charge. 

In the next section we shall discuss stationary and axisymmetric spacetimes, solutions of \eqref{action}. As we shall see, we have complete integrability of the Lewis-Papapetrou system in the same sense as in GR. We will then explicit the Ernst method for this theory and give as a concrete example the construction of the NUT charged BBMB solution. Having done this it will be straightforward to add a cosmological constant, and Maxwell and axionic charges. We will show that the hyperbolic version of Taub-NUT-AdS with conformal scalar is actually a rotating black hole with a well defined horizon. Finally, in the concluding section we will discuss, in light of our findings, the simpler case of hyperbolic Taub-NUT-AdS metric and argue that it is a rotating black hyperboloid membrane in AdS. The black hole curvature singularity, present for $n=0$, is replaced by a chronology violating region covered by inner and outer horizons, in a way that is reminiscent of the three-dimensional BTZ solution \cite{Banados:1992gq}.

%
%
\section{Stationary and axisymmetric spacetimes \label{section2}}

Consider stationary and axisymmetric metrics, solutions of the theory defined by the action \eqref{action} for $\La=0$ and $\al=0$. This means that we are assuming the existence of two commuting Killing vectors $k$ and $m$ such that the former is asymptotically timelike, and the latter  is spacelike and has closed orbits. It is natural to impose the same symmetries to the scalar field $\phi$, that is ${\mc L}_k \phi=\mc L_m\phi = 0$, where $\mc L_X$ denotes the Lie derivative with respect to the fields $X$. In GR, all stationary and axisymmetric metrics can be written  in the Lewis-Papapetrou form, see for example \cite{Straumann:2004fk}. This can be extended to the case of a cosmological constant, but here, given that our metrics are not Einstein metrics, we are not ensured that the Frobenius conditions are still verified for the above gravitational action (\ref{action}). We provide a proof of these conditions in appendix \ref{appendix}. Consequently, using the relations~\eqref{frobenius}, assuming $\phi^2 \neq \frac{3}{4\pi G}$, and skipping some details \cite{Straumann:2004fk}, we can introduce a coordinate system $(t,r,z,\varphi)$ such that any stationary and axisymmetric solution of the theory \eqref{action} can be recast in the Lewis-Papapetrou form,
\BE \d s^2 = - e^{2\la} \lp\d t + A \d\varphi\rp^2 + e^{2(\nu-\la)} (\d r^2 + \d z^2) + \al^2 e^{-2\la}\d\varphi^2 \label{papa0},
\EE
where $\al$, $\la$, $\nu$, $A$ and the scalar field $\phi$ are functions of variables $(r,z)$. When $A=0$, the geometries are static and form the Weyl class of metrics. What is important for our solution-generating purposes is that the form of the field equations approaches that of the equivalent set-up in vacuum GR. Towards this aim, we define $\varepsilon=\pm1$ to be the sign of $\tilde G$, carry out the following redefinitions of the metric functions,
\BE
\be=\varepsilon \lp 1-\frac{4\pi G}{3}\phi^2 \rp \al, \qquad
e^{2\om}=\varepsilon \lp 1-\frac{4\pi G}{3}\phi^2 \rp e^{2\la}, \qquad
e^{\chi}=\varepsilon \lp 1-\frac{4\pi G}{3}\phi^2 \rp e^\nu,
\EE
and rewrite the scalar field in term of a function $\ga$ such that
\BE
\tanh\ga = \lp \sqrt{\frac{4\pi G}{3}} \phi\rp^\varepsilon.
\EE
With these redefinitions, the above metric becomes
\BE \d s^2 = \frac14 \lp e^\ga +\varepsilon e^{-\ga} \rp^2\lb 
-e^{2\om}\lp \d t+A\d \varphi\rp^2 + e^{2(\chi-\om)}\lp \d r^2 + \d z^2 \rp + \be^2 e^{-2\om}\d\varphi^2
\rb . \label{metric-papa}\EE
We see that this metric 
is similar to the one of vacuum GR with the same symmetries -- the well-known Lewis-Papapetrou form (see \cite{Papapetrou:1953zz,Papapetrou:1966zz}) -- but now we have an extra conformal factor in \eqref{metric-papa} which is simply the dimensionless effective gravitational constant $\tilde{G}/G$.
Indeed, introducing complex coordinates $u=\frac{r-iz}2$ and $v=\frac{r+iz}2$, the equations of motion \eqref{EOMg}-\eqref{EOMphi} can be rewritten as the following system of coupled differential equations:
\begin{align}
\be_{,uv} &= 0 \label{papa1} \\
A_{,uv} - \frac12 \lp A_{,u}\frac{\be_{,v}}{\be}+A_{,v}\frac{\be_{,u}}{\be}\rp + 2 A_{,u} \om_{,v} + 2 A_{,v} \om_{,u} &= 0 \label{imaginary}\\
\om_{,uv} + \frac12 \lp \om_{,u}\frac{\be_{,v}}{\be}+\om_{,v}\frac{\be_{,u}}{\be}\rp + \frac{e^{4\om}}{2\be^2}A_{,u} A_{,v} &= 0 \label{real}\\
\chi_{,uv} + \om_{,u}\om_{,v} + 3 \ga_{,u}\ga_{,v} + \frac{e^{4\om}}{4\be^2}A_{,u} A_{,v} &=0 \label{papa4} \\
\ga_{,uv} + \frac12\lp \ga_{,u}\frac{\be_{,v}}{\be}+\ga_{,v}\frac{\be_{,u}}{\be}\rp &= 0 \label{papa5}\\
2\frac{\be_{,u}}{\be}\chi_{,u} -\frac{\be_{,uu}}{\be} &= 2\om_{,u}^2 + 6\ga_{,u}^2 - \frac{e^{4\om}}{2\be^2} A_{,u}^2 \qquad (u \leftrightarrow v) \label{papa6}
\end{align}
Note that equation \eqref{papa4} can be deduced from the others. 

The important result here is the structure of the field equations in that they are quasi-identical to those governing a pure Einstein geometry in standard Lewis-Papapetrou form \cite{Papapetrou:1953zz,Papapetrou:1966zz}.  The only difference emanates from the presence of the field $\ga$ encoding the additional scalar in the action. 

First, note that $\be$ is again a harmonic function and can be set to $\be=r$ without any loss of generality. Then we have a coupled system of equations for the pair $(\om,A)$, which can be treated in numerous ways (see for example \cite{KSMH} and the multitude of references within). Once determined, these fields can be substituted in the non-linear equations \eqref{papa6} governing $\chi$. Here, the only difference with the vacuum case is the extra linear equation \eqref{papa5} that determines $\ga$. This equation is also of the Weyl form \eqref{real}
(in absence of the $A_{,u}A_{,v}$ term), and the resulting field $\ga$ sources the equations $\eqref{papa6}$ for $\be$. In other words, the field equations originating from the action \eqref{action} consist -- for the given symmetries -- are closely related to corresponding Lewis-Papapetrou equations of ordinary GR, as described in \cite{KSMH}. Following Astorino \cite{Astorino:2013xc}, one could have chosen to work in the minimal frame 
and conformally transform for each solution. Here instead, we give the Lewis-Papapetrou form directly in the frame of interest, allowing to construct directly novel solutions using standard GR methods such as those of Ernst or Papapetrou. As we shall see, this has the advantage of preserving some of the GR intuition concerning the relevant sources to use. For a full list of solution generating methods that can be applied to the problem one can consult \cite{KSMH}. Our starting point will be to treat the simpler seed solutions first in their Weyl form, and then to seek the stationary solutions. 

Towards this end we now look into the simpler sub-case of a static and axisymmetric spacetime, which is equivalent to putting $A=0$ in \eqref{metric-papa} and in the field equations \eqref{papa1}-\eqref{papa6}. We choose $\be=r$ by virtue of \eqref{papa1} and it follows that \eqref{real} and \eqref{papa5} are two Laplace equations written in cylindrical coordinates,
\BE \om_{,rr} +\frac1{r}\om_{,r} + \om_{,zz} =0
\qquad\text{and}\qquad
\ga_{,rr} +\frac1{r}\ga_{,r} + \ga_{,zz} =0 .
\EE
In GR, only the Weyl potential $\om$ is present. 
Here, we have two Weyl potentials, similarly to what happens in higher dimensional spacetimes  \cite{Emparan:2001wk}.
Furthermore, these equations are linear, and the Weyl potentials can therefore be superposed. Once $\ga$ and $\om$ are determined, they can be substituted into \eqref{papa6} to determine the function $\chi$ which encodes the non-linearity of Einstein's equations. The Weyl problem is an integrable problem in GR and we thus have shown that this property is also true for the theory \eqref{action}. There are several ways to proceed in order to find solutions \cite{KSMH}; here, we simply outline Weyl's original method, that makes a nice parallel to Newtonian gravity in the spatial dimensions. 

We can interpret the function $\om$ as a solution of Poisson's equation, $\Delta \om = - 4\pi \rho$, with a source term $\rho$ which represents some Newtonian source on the axis of symmetry of the cylindrical space. It turns out that the presence of an event horizon corresponds to taking a localized linear mass distribution -- a \textit{rod} -- located on the axis $r=0$. The gravitational potential takes the standard form
\BE  \om(\vec{r}) = - \int\frac{\rho(\vec{x})}{|\vec{r}-\vec{x}|}\,\mathrm{d}^3{x}.
\EE 
In particular, if we consider a uniform rod on the $r=0$ axis, stretched between $z=z_1$ and $z=z_2$ in cylindrical coordinates, and with density $\si_\om$ per length unit, we find
\BE \om(r,z) = -\si_\om\int_{z_1}^{z_2}\frac{\d\tilde{z}}{\sqrt{r^2+(z-\tilde{z})^2}} = \si_\om\ln\lp\frac{\sqrt{r^2+(z-z_1)^2}-(z-z_1)}{\sqrt{r^2+(z-z_2)^2}-(z-z_2)}\rp \label{distribution}\EE
This is the standard procedure by which a massive rod is shown to correspond to the Schwarzschild solution in GR. 
Note that the field $\gamma$,  representing the scalar field, solves a similar Poisson equation, with its source again localized on the axis of symmetry.

\subsection{Rod structure and the BBMB solution}

As mentioned in the introduction, a notable solution to the equations of motion \eqref{EOMg}-\eqref{EOMphi} is the BBMB solution \cite{BBMB,Bekenstein:1974sf}. It is a static and spherically symmetric configuration, given by the metric
\BE \d s^2 = - (1- m/\rho )^2 \d t^2 + \frac{\d \rho^2}{(1-m/\rho)^2} + \rho^2 \lp \d\theta^2 + \sin^2\theta \d\varphi^2 \rp \label{BBMBg}\EE 
and the following scalar field,
\BE \phi = \sqrt{\frac3{4\pi G}} \frac{m}{\rho-m} . \label{BBMBphi}
\EE
This solution is in many ways special. It is a black hole dressed with a non trivial massless scalar field and secondary hair (note the presence of only one integration constant $m$). It has a well defined geometry, identical to that of the extremal Reissner-Nordstr\"om black hole: the effect of the conformally coupled scalar field on the geometry is indeed to shape it in a similar way to an electric charge at extremality. The scalar field however, unlike the electric field,  explodes at the location of the event horizon (see  \cite{Bekenstein:1975ts} for a discussion on the properties of this solution). Scalar-tensor theories admitting other than ordinary GR solutions are rather rare. This is what makes this black hole so special, and motivates the detailed study of the conformally coupled theory in question, defined by the action \eqref{action}.
%
%
The rod structure of the BBMB solution bears an interesting analogy with ordinary GR. This is manifest if we recast its metric \eqref{BBMBg} in the form \eqref{metric-papa}. For definiteness, let us restrict to the $\rho>2m$ region, where $\tilde G>0$.
To this end, we first trade the radial coordinate $\rho$ for an auxiliary coordinate $u = \ln\lp\rho/m-1\rp$, and then we switch to cylindrical coordinates $(t,r,z,\varphi)$ throught the coordinate transformation
\BE
z = 2m \cosh u \cos\theta \qquad\text{and}\qquad r = 2m \sinh u \sin\theta.
\label{zr}\EE
As a result, the BBMB metric becomes of the form \eqref{metric-papa} with
\BE e^{4\omega} = e^{-4\ga} = \frac{\sqrt{r^2+(z+2m)^2}-(z+2m)}{\sqrt{r^2+(z-2m)^2}-(z-2m)} \label{weyl2}\EE
and
\BE e^{2\chi} = \frac{\sqrt{r^2+(z-2m)^2}\sqrt{r^2+(z+2m)^2}+(z-2m)(z+2m)+r^2}{2\sqrt{r^2+(z-2m)^2}\sqrt{r^2+(z+2m)^2}}. \label{weyl3}\EE
Comparing with \eqref{distribution}, we see that $\om$ and $\ga$ are solutions of the Laplace equation with a source located at $r=0$ between $z=-2m$ and $z=2m$ with a density $\si_\om=1/4$ and $\si_\ga=-1/4$ respectively. 

\begin{figure}[h]
\begin{picture}(0,0)
\put(15,25){\rule{0.8\textwidth}{2pt}}
\put(65,24){\rule{0.2\textwidth}{7pt}}
\put(58,28){-$2m$}
\put(99,28){$2m$}
\put(73,20){$\si_\om=1/4$}
\put(149,25){$\om$}

\put(15,5){\rule{0.8\textwidth}{2pt}}
\put(65,4){\rule{0.2\textwidth}{7pt}}
\put(58,8){-$2m$}
\put(99,8){$2m$}
\put(73,0){$\si_\ga=-1/4$}
\put(149,5){$\ga$}
\end{picture}
\caption{Rod structure of the BBMB solution}
\label{rod-BBMB}
\end{figure}
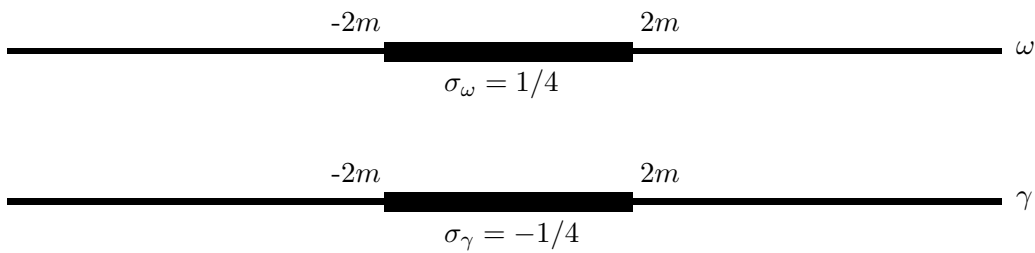

We can specify that these distributions are placed at $\rho=2m$ in the original coordinates $(t,\rho,\theta,\varphi)$ used in \eqref{BBMBg} and \eqref{BBMBphi}. On the other hand, when $m<\rho<2m$, where $\phi^2>\frac{3}{4\pi G}$, we set $u=-\ln(\rho/m-1)$. Note that Weyl coordinates require two differing patches of the same radial $\rho$ coordinate \eqref{BBMBg}. Apart from this fact we find the same rod structure as before summed up in figure \ref{rod-BBMB}. Concerning the function $\chi$, we verify in both cases that $\lim\limits_{r \to 0} \chi(r,z) = 0$ for any $z<-2m$ and $z>2m$ according to \eqref{weyl3}, that there is no conical singularity on the axis.

We can point out that all spherically symmetric and static solutions of \eqref{action}, given in \cite{Barcelo:1999hq} and parametrized by three constants $(m,\ep,\delta)$, can be generated using this rod structure with the following densities $\si_\om = \frac{\cos\ep}{2}$ and $\si_\ga=-\frac{\sin\ep}{2\sqrt{3}}$ and the same lengths $4m$ (see \cite{Bardoux:2012nx} for the details). The third parameter $\delta$, is a constant that we can always add to the Weyl potential $\ga$ in addition to the logarithmic term associated to the rod while keeping the same asymptotic properties. In complete analogy to GR, the C-metric version of the BBMB solution \cite{Charmousis:2009cm,Anabalon:2009qt} can be obtained by adding a semi-infinite $\om$-rod of density $1/2$ to the previous rod structure of the BBMB solution (see fig.\ \ref{rod-BBMB}). 

\vspace{4cm}

\subsection{The Ernst method}

So much for the Weyl formalism. When treating the full stationary and axisymmetric problem (\ref{papa1})-(\ref{papa6}) one can make use of a very elegant method developed by Ernst, \cite{Ernst:1967wx} which turns the coupled PDE's (\ref{imaginary})-(\ref{real}) for the fields $(\om, A)$ into a single PDE of a complex variable. Here, the form of (\ref{imaginary})-(\ref{real}) is identical to the one in pure GR so we can just iterate the method, by introducing  an auxiliary  field $\Om(r,z)$ such that
\BE \lp A_{,r}, A_{,z} \rp = \lp \be e^{-4\om} \Om_{,z}, -\beta e^{-4\om} \Om_{,r}\rp .\EE
Then the complex function $ \mc E = e^{2\om} + i \Om $, called the \textit{Ernst potential}, verifies the \textit{Ernst equation}
\BE \frac1{\be}\vec{\n}.\lp\be\vec{\n}\mc E\rp = \frac{\vec{\n}\mc E . \vec{\n}\mc E}{\mathrm{Re}\lp\mc E\rp},\label{ernst1}\EE
where $\vec{\n}=\vec{u}_r \p_r + \vec{u}_z \p_z$ is the gradient in the euclidean $r$-$z$ plane. The real and imaginary parts of this equation correspond to \eqref{real} and \eqref{imaginary} respectively. Redefining the Ernst potential with a new complex function $\xi$ so that
\BE \mc E = \frac{\xi-1}{\xi+1} , \label{xi}\EE 
equation \eqref{ernst1} takes the form \cite{Ernst:1967wx}
\BE
\lp\xi\xi^*-1 \rp\frac1{\be}\vec{\n}.\lp\be\vec{\n}\xi\rp = 2\xi^* \vec{\n}\xi . \vec{\n}\xi.
\label{ernst2}\EE
This equation enjoys a $U(1)$ invariance: if $\xi$ is a solution, $e^{i\la}\xi$ is also a solution, for any real $\la$. According to \eqref{papa1}, we adopt henceforth the choice $\be=r$. It is then useful to introduce prolate spheroidal coordinates $(x,y)$ defined by
\BE
r = \si\sqrt{x^2-1}\sqrt{1-y^2} \qquad\text{and}\qquad z = \si x y \label{xy},
\EE
where $x$ and $y$ are dimensionless coordinates, and $\si$ is a constant with the dimension of length. In particular, we get
\BE \d r^2 + \d z^2 = \si^2 (x^2-y^2) \lp \frac{\d x^2}{x^2-1} + \frac{\d y^2}{1-y^2} \rp.\EE
Knowing $\xi$, we can infer the real and imaginary parts of the Ernst potentials $\lp e^{2\om},\Om \rp$, and determine the function $A$ using
\BE A_{,x} = \si(1-y^2)e^{-4\om}\Om_{,y} \qquad\text{and}\qquad A_{,y} = \si(1-x^2)e^{-4\om}\Om_{,x} \label{rotation}\EE
and finally deduce the function $\chi$ with
\begin{align}
\chi_{,x} &= \frac{1-y^2}{4(x^2-y^2)}e^{-4\om}
\left[x(x^2-1)B_{xx} + x(y^2-1)B_{yy} + 2y(1-x^2)B_{xy}\right],
\label{chix}\\
\chi_{,y} &= \frac{x^2-1}{4(x^2-y^2)}e^{-4\om}
\left[y(x^2-1)B_{xx} + y(y^2-1)B_{yy} +2x(1-y^2)B_{xy}\right],
\label{chiy}
\end{align}
where
\BE B_{ij} = \lp e^{2\om}\rp_{,i} \lp e^{2\om}\rp_{,j} + \Om_{,i} \Om_{,j} + 3e^{4(\om-\ga)}\lp e^{2\ga}\rp_{,i} \lp e^{2\ga}\rp_{,j} .\EE
Summarizing, given the potentials $(\ga,\xi)$, we have shown that following step by step Ernst's original method, we can determine completely a stationary and axisymmetric solution of \eqref{action}. 
A treatment of this problem in vacuum, higher dimensional GR can be found for example in \cite{Harmark:2005vn}. The Ernst equation in presence of a cosmological constant was studied in \cite{Charmousis:2006fx,Astorino:2012zm}, and used to construct a Melvin solution with a cosmological constant \cite{Astorino:2012zm}.

\subsection{Example: including Taub-NUT charge}

As a concrete example of the above method, we now construct the Taub-NUT extension of the BBMB solution. A similar construction is given for GR in 
\cite{reina1975nut}  where one starts from Schwarzschild as the Ernst potential for the seed solution (see also \cite{gautreau1972generating} for a construction based on the Papapetrou method). Here our seed solution is quite naturally BBMB, for which $\si=2m$. According to \eqref{zr} and \eqref{xy}, we transform  to prolate spheroidal coordinates $(x,y)$ for the BBMB solution by setting $x=\cosh u $ and $y=\cos\theta$. In particular, the seed Ernst potential, which is real since the solution is static, is given by
\BE
e^{2\om} = \lp\frac{x-1}{x+1}\rp^{1/2} = \frac{x\pm\sqrt{x^2-1}-1}{x\pm\sqrt{x^2-1}+1} ,
\EE
by virtue of \eqref{weyl2}. As a result, $\xi_0$ defined in \eqref{xi} is $\xi_0=x\pm\sqrt{x^2-1}$. Then, according to \eqref{ernst2}, $\xi=e^{i\la}\xi_0$ is also a solution of \eqref{ernst2} and the resulting complex Ernst potential reads,
\BE \mc E = \frac{\sqrt{x^2-1}}{x+\cos\la} + i \frac{\sin\la}{x+\cos\la} . \EE
In turn, equations \eqref{rotation} give,
\BE A = 2m y \sin\la .  \EE
As for the field $\ga$ representing the scalar field, we have a choice. One can keep the same Weyl potential as for BBMB (see figure \ref{rod-BBMB}) or, again, take something more general. It turns out that keeping the same scalar field as for the BBMB solution, that is
\BE e^{2\ga} = \lp \frac{x+1}{x-1} \rp^{1/2} \EE
is the relevant choice. This evades extra singularities since  \eqref{chix} and \eqref{chiy} give
\BE e^{2\chi} = \frac{x^2-1}{x^2-y^2} , \EE
and this is the same field as the one for the BBMB solution since it is independent of the phase $\la$. In $(t,\rho,\theta,\varphi)$ coordinates, we obtain,
\BE \d s^2 = - \frac{(\rho-m)^2}{f(\rho)} \lp \d t + 2m\sin\la\cos\theta\d\varphi \rp^2+ \frac{f(\rho)}{(\rho-m)^2}\d\rho^2 + f(\rho)\lp\d\theta^2+\sin^2\theta\d\varphi^2\rp \EE
where
\BE f(\rho) = \rho^2 + 2m(\cos\la-1)\rho + 2m^2(1-\cos\la) .\EE
Finally, shifting $\rho\rightarrow \varrho = \rho + m \lp \cos\la-1 \rp $, introducing the NUT parameter $n=m\sin\la$, and redefining the mass parameter through $\mu=\sqrt{m^2-n^2}$, we recast the solution into a more familiar form,
\begin{align}
\d s^2 &= - \frac{(\varrho-\mu)^2}{\varrho^2+n^2} \lp \d t + 2n\cos\theta\,\d\varphi \rp^2 + \frac{\varrho^2+n^2}{(\varrho -\mu)^2} \d \varrho^2 + \lp \varrho^2+n^2 \rp \lp\d\theta^2+\sin^2\theta\,\d\varphi^2\rp \label{NUT-BBMB-g} \\
\phi &= \sqrt{\frac{3}{4\pi G}} \frac{\sqrt{\mu^2+n^2}}{\varrho-\mu} . \label{NUT-BBMB-phi}
\end{align}

Quite naturally, the BBMB solution is recovered when the NUT parameter vanishes. Furthermore, as for BBMB, the metric geometry of \eqref{NUT-BBMB-g} is identical to that of the extremal charged Taub-NUT solution first found in \cite{Brill:1964fk}. The Taub region of the metric, which is the non-stationary, timelike 
region situated in between the inner and outer horizons of the bald metric, is now absent given the extremal nature of the horizon. The solution is only locally asymptotically flat as asymptotic infinity is deformed by the NUT charge. In fact, although the Riemann curvature decays at asymptotic infinity at the same rate as in the Schwarzschild solution or BBMB, the Taub-NUT version is not a solution of a $1/\varrho$ Newtonian  deformation about flat spacetime \cite{misner1963flatter}. The scalar field explodes at the horizon location $\varrho=\mu$ as for the BMBB metric. Despite this singularity the metric \eqref{NUT-BBMB-g} is regular for all $\varrho$. In fact, $\varrho=0$ is not even a singular point for the metric! The main pathology of Taub-NUT solutions is the presence of the Misner string either at $\theta=0$ or $\theta=\pi$ rendering the 1-form $\d\varphi$ ill-defined at either of these points. In order to get rid of the North pole singularity at $\theta=0$, we can define a new time coordinate $t'=t+2n\varphi$. For the south pole at $\theta=\pi$ we take $t''=t-2n\varphi$. Given the periodicity of $\varphi$ by $2\pi$ we now have to identify $t'$ and $t''$ modulo $8\pi n$. Hence, the only way out is to impose periodicity of the time coordinate by $t\rightarrow t+8\pi n$. The resulting one form in $\d\varphi$ is well defined but the price to pay are CTCs in the Lorentzian metric{\footnote{Note that hypersufaces orthogonal to $t=\mathrm{const}$ are space like only for sufficiently small $\theta$. In other words the causality inconsistencies of Lorentzian Taub NUT are not only a result of imposing time periodicity \cite{misner1963flatter}, they are there in the metric element from the start.}} \eqref{NUT-BBMB-phi}. We are therefore led to consider as usual the regularity of the euclidean metric. For that, set $t\rightarrow i\tau$ and $n\rightarrow in$ on \eqref{NUT-BBMB-g}-\eqref{NUT-BBMB-phi}
\begin{align}
\d s^2 &= \frac{(\varrho-\mu)^2}{\varrho^2-n^2} \lp \d \tau + 2n\cos\theta\d\varphi \rp^2 + \frac{\varrho^2-n^2}{(\varrho -\mu)^2} \d \varrho^2 + \lp \varrho^2-n^2 \rp \lp\d\theta^2+\sin^2\theta\,\d\varphi^2\rp \\
\phi &= \sqrt{\frac{3}{4\pi G}} \frac{\sqrt{\mu^2-n^2}}{\varrho-\mu}. 
\end{align}
As before, to have no Misner string, the periodicity of $\tau$ is fixed at the value $8\pi n$. But now $\tau$ is just parametrising the circle which is fibered over the 2-sphere of coordinates $\theta$ and $\phi$. Following the classification of gravitational instantons given in \cite{Gibbons:1979xm} we now have a curvature singularity at $\rho=n$ and hence no nut solutions can be defined unless $\mu=n$. This is the case where the scalar field is set to zero and in fact we go back to the Ricci flat nut instanton. When $\mu>n$ the singularity at $\rho=n$ is never reached since space closes off smoothly at $\varrho=\mu$ due to the extremal nature of the Lorentzian horizon. In this case the $U(1)$ isometry generated by the Killing vector $\p_\tau$  has a two-dimensional fixed point set, a bolt, as does Euclidean Schwarzschild, given by $g_{\tau\tau}=0$. The fixed point, $\varrho=\mu$, set are the 2-spheres. Finally, due to the extremal nature of the horizon there is no conical singularity (once the periodicity of $\tau$ is imposed). We therefore have a regular metric as long as $\mu \geq n$. The solution is continuously related to the Ricci flat nut solution at $\mu=n$. In the presence of NUT charge therefore the transition in between the BBMB and Schwarzschild family is smooth. The scalar explodes at $\rho=\mu$ but we can remedy this including a cosmological constant. This is what we will consider next. 

\section{Including matter and a cosmological constant \label{topo}}

We now turn on the cosmological constant $\La$ and the self-interaction potential in $\eqref{action}$. We will also add in turn, the electromagnetic (EM) interaction,
\BE S_\mathrm{em}[g_{ab}, A_a]=-\frac1{16\pi G} \int_{\mc M} \sqrt{-g} F_{ab}F^{ab}\d^4 x . \label{action1}\EE
and an axionic interaction,
\BE S_\mathrm{ax}=- \frac12 \sum_{i=1}^2 \int_\mc M \lp 1-\frac{4\pi G}3 \phi^2\rp^{-1} \mc H^{(i)} \w \star \mc H^{(i)} . \label{action2}\EE
We add here two exact three-forms $\mc H^{(i)} = \frac1{3!} H^{(i)}_{abc} \d x^a \w \d x^b \w \d x^c$, $(i = 1, 2)$, originating from two Kalb-Ramond potentials $\mc B^{(i)}$ such that $\mc H^{(i)} = \d \mc B^{(i)}$. Although the EM interaction is minimally coupled in four dimensions for the conformal frame the axions are non-minimally coupled to the scalar field $\phi$. It is perhaps more illuminating to write this action  with a minimal coupling for the scalar field. Performing the following conformal transformation,
\BE
\label{conformal}
\tilde{g}_{ab}=\lp 1 - \frac{4\pi G}{3}\phi^2 \rp g_{ab}
\EE
and a  redefinition for the scalar field $\Psi = \sqrt{\frac{3}{4\pi G}}\,\artanh\lp \sqrt{\frac{4\pi G}{3}}\phi\rp$, the full action \eqref{action2} becomes
\BE S = \int_{\mc M}  \sqrt{-\tilde{g}}\lb\frac{\tilde{R}}{16\pi G}   - \frac{1}{2} \sum_{i=1}^2  \frac{1}{3!} H_{abc}^{(i)} H^{(i)abc}  - \frac{F_{ab}F^{ab}}{16\pi G} - \frac{1}{2}\p_a \Psi \p^a \Psi - U(\Psi)  \rb \d^4x \EE
where the scalar potential is given by
\BE U(\Psi) = \frac{\La}{8\pi G}\left[ \cosh^4\left(\sqrt{\frac{4\pi G}{3}}\Psi\right) + \frac{9\al}{2\pi\La G} \sinh^4\left(\sqrt{\frac{4\pi G}{3}}\Psi\right) \right] .\EE

Let us turn now to the equations of motion in the conformally coupled frame. Variation of the full action \eqref{action}, \eqref{action1} and \eqref{action2} with respect to the metric gives
\begin{multline}\label{EOM2g}
 \lp 1-\frac{4\pi G }{3}\phi^2\rp  G_{ab}  + \La g_{ab}  = 8\pi G \lp 1-\frac{4\pi G }{3}\phi^2\rp^{-1} \sum_{i=1}^2 \lp\frac{1}{2}H_{acd}^{(i)} H_b^{(i)cd} - \frac{1}{12}g_{ab} H_{cde}^{(i)} H^{(i)cde} \rp \\
 + 2 \lp F_{ac}F_b^{\ c} - \frac{1}{4}g_{ab}F_{cd} F^{cd} \rp  + 8\pi G \lp \p_a \phi \p_b \phi - \frac{1}{2}g_{ab} \p_c \phi \p^c \phi \rp \\
 + \frac{4\pi G}{3}\lp g_{ab} \Box - \n_a \n_b \rp\phi^2 - 8\pi G \al g_{ab} \phi^4 .
\end{multline}
Then varying with respect to $\phi$, $\mc A$ and $\mc B^{(i)}$, we obtain
\BE\Box\phi = \frac{R}{6}\phi + \frac{4\pi G}{3} \phi \left( 1 - \frac{4\pi G}{3} \phi^2 \right)^{-2} \sum_{i=1}^2  \frac{1}{3!} H_{abc}^{(i)} H^{(i)abc} + 4\al \phi^3,
\label{EOM2phi}\EE
\BE \n_a F^{ab} = 0  \qquad\text{and}\qquad  \n_a \lb \lp 1 - \frac{4\pi G}{3} \phi^2 \rp^{-1} H^{(i)abc} \rb =0\EE
respectively. An important property of the field equations stemming from the conformal coupling is the following: taking the trace of the metric equations \eqref{EOM2g} and replacing it in the equation of motion for the scalar field \eqref{EOM2phi} gives
\BE \Box\phi = \frac{2}{3}\La\phi + 4\al \phi^3. \EE
This is precisely the equation of motion \eqref{EOMphi2} which emanates from the theory \eqref{action2} in the absence of matter fields. Hence we can try to find a solution related to a hairy solution of the non-axionic theory. Finally, given the equation of motion for the axionic fields, it will be useful to introduce pseudo-scalar fields $\chi^{(i)}$ which are dual to the axionic fields in the following sense
\BE \mc H^{(i)} = \lp 1-\frac{4\pi G}{3}\phi^2 \rp \star \d \chi^{(i)} . \label{scalar}\EE

We expect that the presence of the cosmological constant will permit to hide the divergence of the scalar field \cite{Martinez:2005di}. Let us first switch off the axionic fields for the solutions are completely different in nature for the latter case.

\subsection{Rotating black holes with a cosmological constant}

In this section we consider \eqref{action} including a cosmological constant and an EM-field \eqref{action1}.
Given the Taub-NUT-BBMB construction \eqref{NUT-BBMB-g}-\eqref{NUT-BBMB-phi}, it is not difficult to guess the solution including  the additional terms. The equations of motion \eqref{EOMg}-\eqref{EOMphi} admit the following solution,
\BE \d s^2= - V(\varrho) \lp \d t + \mc B \rp^2 + \frac{\d \varrho^2}{V(\varrho)}  + (\varrho^2+n^2) \d \si^2_{(k)}\label{solution-g}\EE
with
\BE V(\varrho) = - \frac{\La}3 (\varrho^2+n^2) + \lp k-\frac43 n^2 \La \rp \frac{(\varrho-\mu)^2}{\varrho^2+n^2} \label{solution-V} \EE
and
\BE \mc B =
\left\{ \begin{aligned} 
2n\cos\theta\d\varphi \qquad &\text{when} \qquad \d \si^2_{(k=1)} = \d \theta^2 + \sin^2\theta \d \varphi^2  \\
n\theta^2\d\varphi \qquad &\text{when} \qquad \d \si^2_{(k=0)} = \d \theta^2 + \theta^2 \d \varphi^2 \\
2n\cosh\theta\d\varphi \qquad &\text{when} \qquad \d \si^2_{(k=-1)} = \d \theta^2 + \sinh^2\theta \d \varphi^2  \\
\end{aligned} \right. \label{solution-B}
\EE
The scalar field and the electromagnetic potential read  
\BE \phi = \sqrt{-\frac{\La}{6\al}} \frac{\sqrt{\mu^2+n^2}}{\varrho-\mu} \label{phi}\EE
and
\BE \mc A = \frac{q\varrho}{\varrho^2+n^2}\lp \d t + \mc B \rp \label{A}\EE
respectively. Thus  $\al \La<0$  to get a real scalar field and it is interesting to note that the Faraday field carries both an electric and a magnetic component. Finally, the integration constants $(n,\mu,q)$ satisfy the following constraint
\BE q^2 = \lp \mu^2+n^2 \rp \lp k-\frac43 n^2 \La \rp \lp 1+\frac{2\pi \La G}{9\al} \rp . \label{constraint}\EE
For a vanishing NUT parameter, and for $k=1,-1$ we recover the topological black hole solutions of \cite{Martinez:2005di}. Secondly, when $\La$ and $\al$ go to zero so that $-\frac{\La}{6\al}$ tends to the value $\frac3{4\pi G}$, we obtain the Taub-NUT-BBMB solution \eqref{NUT-BBMB-g}-\eqref{NUT-BBMB-phi} for $k=1$. 

In order for the metric to be regular we need to evade the Misner strings whenever these are present. For spherical sections, $k=1$, these are unavoidable given the presence of non zero NUT charge. As a result CTCs will appear in the periodic orbits of time just like for usual NUT spacetime. One can consider a Euclidean signature metric as in the previous section. When $k=0$ however, the $t$-fibration on the plane is trivial and thus no Misner strings are present \cite{Chamblin:1998pz}. For $\Lambda<0$ however, we now note that for large enough $\varrho$ and $\theta$ the Killing vector $\partial \varphi$ becomes time like. Given that $\varphi$ is periodic this means that we have closed timelike curves for $\varphi=constant$ and large enough $\varrho$ and $\theta$. The situation is similar to the one encountered in G\"{o}del spacetime, \cite{Hawking:1973uf}. For $\Lambda>0$ (and $k=0$), the nature of the solution completely changes, for then, $\varrho$ is actually everywhere timelike and the solution is of cosmological nature, homogeneous and non-isotropic. The $\varrho$ coordinate sweeps the whole real line and the solution is completely free of coordinate or curvature singularities. The scalar field explodes at $\varrho=\mu$. At large $\varrho$ spacetime  locally asymptotes de-Sitter. We are thus in the Taub region of Taub-NUT, although this region is now covering the whole of spacetime. It is the hyperbolic case $k=-1$ that gives the most interesting geometry, a rotating black hole solution. Indeed Misner strings are not present and then it is easy to note, defining first $T=t+2n\varphi$ that $g_{\varphi\varphi}>0$ for all $\rho$ and $\theta$ as long as $4n^2 l^2<1$. Therefore there are no CTCs whatsoever. This, to our knowledge, is a novel property for Taub NUT metrics and is obtained due to the presence of the conformally coupled scalar field. This spacetime geometry is very peculiar and interesting for the radial coordinate $\varrho$ can be ``extended'' to range from $-\infty$ to $+\infty$ starting and ending at a hyperbolic slicing of AdS. In this way the metric has up to four Killing horizons, but no spacetime singularity whatsoever! I As a result we can take $\mu>0$ without any loss of generality.
The up to four Killing horizons are located at the zero's of $V(\varrho)$ and for $\varrho\in\mathbb R$,
\begin{align}
\varrho_+&= \frac{\bar{l}}2\lp 1+\sqrt{1-4\frac{n^2}{\bar{l}^2}-4\frac{\mu}{\bar{l}}} \rp\\
\varrho_-&=\frac{\bar{l}}2\lp 1-\sqrt{1-4\frac{n^2}{\bar{l}^2}-4\frac{\mu}{\bar{l}}} \rp\\
\varrho_{++}&=\frac{\bar{l}}2\lp -1+\sqrt{1-4\frac{n^2}{\bar{l}^2}+4\frac{\mu}{\bar{l}}} \rp \\
\varrho_{--}&=\frac{\bar{l}}2\lp -1-\sqrt{1-4\frac{n^2}{\bar{l}^2}+4\frac{\mu}{\bar{l}}} \rp
\end{align}
according to the values of $\mu$, $n$ et $\bar{l}=\sqrt{l^2-4n^2}$. The metric has no other singular points whatsoever. The case of $n=0$ is a black hole  and has been studied in \cite{Martinez:2005di}. Note that the solution presented here is, for $n\neq 0$ regularizing the $\Lambda-$BBMB solution which has a spacetime curvature singularity at $\varrho=0$. We have thus a rotating solution which at the price of changing the asymptotic structure regularizes its static counterpart metric. As often the Taub NUT family of metrics presents a counterexample to our usual intuitive understanding of black hole solutions. In fact, we use the term black hole since there is a singularity in the scalar field at $\varrho_0=\mu$. We have that $V(\varrho_0)=\frac{\mu^2+n^2}{l^2}>0$. Hence in order for the scalar singularity to be hidden we need to be in the relevant positive spacetime region, $\varrho_{--}\leq 0\leq \varrho_{++}\leq \mu\leq\varrho_-\leq \varrho_{ext}\leq \varrho_+$ which is verified for $0\leq \mu \leq \frac{\bar{l}}{4}-\frac{n^2}{\bar{l}}$.  We have noted as $\varrho_\mathrm{ext}$ the case $\varrho_-=\varrho_+$ saturating the above bound. It is maybe more relevant from the point of view of singularity theorems to study this metric in the Einstein frame{\footnote{Given that our solution does not couple to matter the choice of physical frame can be in either of frames.}}. If we make a conformal transformation to the Einstein frame,\eqref{conformal},  there is a genuine curvature singularity where the conformal transformation is singular{\footnote{For a discussion see for example \cite{Bardoux:2012tr} and note also the change of sign in the the effective Newton's constant\eqref{Gtilde}}}, $\varrho_1=\mu+\eta \sqrt{\mu^2+n^2}$, where we note for convenience $\eta= -\frac{2\pi \La G}{9\al}\geq 1$. In this case in order for $\varrho_-<\varrho_1< \varrho_+$ we must have,
\BE
\frac{-2\eta^2-(1+\eta^2)\sqrt{\eta^2-\frac{n^2}{\bar{l}^2}(1-\eta^2)^2}}{(1-\eta^2)^2}<\frac{\mu}{\bar{l}}<\frac{-2\eta^2+(1+\eta^2)\sqrt{\eta^2-\frac{n^2}{\bar{l}^2}(1-\eta^2)^2}}{(1-\eta^2)^2}
\EE
which in the Einstein frame guarantees a black hole. For hyperbolic sections, but $\Lambda>0$ we have a similar cosmological metric as for the planar case with $\Lambda>0$.
The maximal NUT charge is attained  when $\frac43 n^2 \La = k$. Given the constraint \eqref{constraint} the electromagnetic field vanishes but not the scalar field which is again singular at $\varrho=\mu$. The spacetime geometry is that of de-Sitter and anti de-Sitter for the relevant sign of $\Lambda$.

\subsection{Bald and hairy solutions with axionic fields\label{shaping}}

We now switch on the axionic fields \eqref{action2}. 
Let us start by presenting the bald solution, that is with a vanishing scalar field, $\phi=0$. For a negative cosmological constant, $\La=-3/l^2$, the theory \eqref{action2} admits the following solution,
\BE \d s^2= - V(\varrho) \lb \d t + n \lp x \d y - y \d x \rp \rb^2 + \frac{\d \varrho^2}{V(\varrho)}  + (\varrho^2+n^2) \lp \d x^2 + \d y^2 \rp  \label{axionicTNRN} \EE
with lapse function
\BE V(\varrho)=\frac1{\varrho^2+n^2}\lb -2m\varrho + \frac1{l^2} \lp \varrho^4+6n^2\varrho^2-3n^4\rp + p^2\lp n^2-\varrho^2\rp + q^2\rb \EE
and the following electromagnetic potential
\BE \mc A = \frac{q\varrho}{\varrho^2+n^2}\lb \d t + n \lp x \d y - y \d x \rp \rb.\EE
For $p=1$, we notice that the lapse function is identical to the Taub-NUT AdS-Reissner-Nordstr\"om solution with a hyperbolic base manifold \cite{Chamblin:1998pz}.  Here, however, we have a flat horizon induced by the presence of two homogeneously distributed 3-forms $\mc H^{(i)}$. These axionic fields are generated by the following scalar fields, defined in \eqref{scalar},
\BE \chi^{(1)}=\frac{p}{\sqrt{8\pi G}}x \qquad\text{and}\qquad \chi^{(2)} = \frac{p}{\sqrt{8\pi G}} y. \EE
To construct explicitly the corresponding axionic fields, we use \eqref{scalar} to obtain,
\BE
\mc H^{(1)} = \frac{p}{\sqrt{8\pi G}} \lp -\d t  + n  y \d x \rp \w \d \varrho \w \d y ,
\EE 
and 
\BE \mc H^{(2)} = \frac{p}{\sqrt{8\pi G}} \lp \d t + n x \d y \rp \w \d \varrho \w \d x .\EE 

First, for $p=0$, we get the Taub-NUT Reissner-Nordstr\"om solution with a flat base manifold briefly discussed in the euclidean time in \cite{Chamblin:1998pz} without the electromagnetic interaction. Secondly, 
for $n=0$, this solution is the axionic planar charged black hole of \cite{Bardoux:2012aw}. Turning on the parameter $n$ makes the metric non-static, and the resulting black hole rotates on the plane of its horizon. Indeed the fibration is trivial and consequently there is no Misner string singularities. There is no period associated to the coordinate time $t$. However, it is easy to note that using polar coordinates $(\theta,\varphi)$ defined by $(x,y)=(\theta\cos\varphi,\theta\sin\varphi)$ that the  norm $g_{\varphi\varphi}$ changes sign for large enough $\varrho$ and $\theta$. This means that we will always have closed timelike curves for in the $2\pi$ periodic $\phi$ orbits. In fact these CTCs can appear arbitrarily close to the outer horizon $\varrho=\varrho_h$ for large enough radial horizon coordinate $\theta$. In fact for any constant $\varrho_0>\varrho_h$ once $\theta_1^2>\theta_0^2=\frac{\rho_0^2+n^2}{V(\rho_0)n^2}$ we have CTCs in constant $(\theta_1,\varrho_0)$ orbits of $\varphi$. Hence these metrics are pathological.

The hairy version is not far better. For a negative cosmological constant, $\La=-3/l^2$, we have,
\BE \d s^2= - V(\varrho) \lb \d t + n(x\d y - y \d x)\rb^2 + \frac{\d \varrho^2}{V(\varrho)}  + (\varrho^2+n^2) \lp \d x^2 + \d y^2 \rp\EE
with
\BE V(\varrho) = \frac{\varrho^2+n^2}{l^2} + \lp \frac{4n^2}{l^2}-p^2 \rp \frac{(\varrho-\mu)^2}{\varrho^2+n^2} \EE
and the following electromagnetic potential and scalar field
\BE \mc A = \frac{q\varrho}{\varrho^2+n^2}\lb \d t + n \lp x \d y - y \d x \rp \rb \quad\text{and}\quad \phi = \frac1{\sqrt{2\al l^2}} \frac{\sqrt{\mu^2+n^2}}{\varrho-\mu} \EE
respectively. Using polar coordinates $(\theta,\varphi)$ defined by $(x,y)=(\theta\cos\varphi,\theta\sin\varphi)$, we notice that for $p=1$ the metric is that of the charged Taub-NUT-BBMB solution \eqref{solution-g}-\eqref{solution-V} with a hyperbolic base manifold corresponding to $k=-1$ in \eqref{solution-B} but here we have a flat horizon induced by the presence of two homogeneously distributed 3-forms $\mc H^{(i)}$ as usual. These axionic fields are generated by the following scalar fields, defined in \eqref{scalar},
\BE \chi^{(1)}=\frac{p}{\sqrt{8\pi G}}x\qquad \chi^{(2)}=\frac{p}{\sqrt{8\pi G}}y ,\EE
and as derived in the previous subsection, it is straightforward to show that the axionic fields take the following form
\BE \mc H^{(1)} = \frac{p}{\sqrt{8\pi G}} \lp 1-\frac{4\pi G}{3}\phi^2 \rp \lp -\d t + n y \d x \rp \w \d \varrho \w \d y ,\EE 
\BE \mc H^{(2)} = \frac{p}{\sqrt{8\pi G}} \lp 1-\frac{4\pi G}{3}\phi^2  \rp \lp \d t + n x \d y \rp \w \d \varrho \w \d x .\EE 
Finally, the integration constants $(n,\mu,q,p)$ must satisfy the following constraint
\BE q^2 = \lp \mu^2+n^2 \rp \lp \frac{4n^2}{l^2}-p^2 \rp \lp 1-\frac{2\pi G}{3\al l^2} \rp .\EE

First, for $p=0$, we get the charged Taub-NUT-BBMB solution \eqref{solution-g}-\eqref{solution-V} with a flat base manifold corresponding to $k=0$ in \eqref{solution-B} analysed in the previous section. Secondly, for $n=0$, this solution is the axionic planar charged black hole with a conformal hair of \cite{Bardoux:2012tr}. Consequently, the solution presented here is the generalization of that solution with the introduction of a rotating parameter $n$. Again this solution has the same CTC pathologies as its bald counterpart.

\section{Conclusions: revisiting the hyperbolic Taub-NUT-AdS metric}
In this article, we have considered scalar-tensor theories of a particular nature: the scalar field is conformally coupled to the four dimensional Einstein-Hilbert action and although the full action is not conformally invariant the theory has nice integrability properties due to the partial conformal invariance. We have shown that the Lewis-Papapetrou metrics and the Weyl problem are integrable in the same way as for vacuum General Relativity \cite{KSMH}. The field equations are augmented by an an additional Weyl potential associated to the conformally coupled scalar field. As such, solution generating methods can be employed in more or less the same manner as for GR, and solutions of the action \eqref{action} can thus be obtained with relative ease. Using the Ernst method \cite{Ernst:1967wx}, we then derived the generic family of Taub-NUT solutions for this theory. This procedure was first developed by Reina and Treves to obtain the relevant Taub-NUT metrics in General Relativity \cite{reina1975nut}. We then derived topological versions by introducing a cosmological constant. The hyperbolic version turns out to be free of closed timelike curves and represents a rotating black hole in the Einstein frame. The solution in the conformal frame is totally free of curvature singularities and the scalar field is singular only in a region hidden behind an event horizon. We then found the bald and hairy axionic solutions, but those are plagued by CTCs which can be even arbitrary close to the horizon. An interesting and yet unresolved problem would be to determine the Kerr family of the action \eqref{action}. Unfortunately here, a direct application of the Ernst method to conformally coupled scalars fails, primarily because the Weyl potential for the BBMB solution is different from that of Schwarzschild black hole, and as a result spheroidal coordinates do not seem to be quite adapted to the problem. This remains an open problem.

Given the hindsight from our research in Taub-NUT metrics, the case of hyperbolic slicing stands out as the most interesting. Indeed it is in this case that the negatively curved horizon permits to do away with the CTC pathology of Taub-NUT. It is interesting therefore to dwell a bit on the simplest of cases, namely that of Einstein-Hilbert with a negative cosmological constant. In this theory, the relevant hyperbolic Taub-NUT-AdS solution can be conveniently written as  \cite{Chamblin:1998pz},
\begin{align}
\d s^2 &= - f(\rho) \lp \d t+ 4n\sinh^2\lp\frac{\theta}{2}\rp\d\varphi \rp^2 + \frac{\d \rho^2}{f(\rho)} + \lp \rho^2+n^2 \rp \lp\d\theta^2+\sinh^2\theta\,\d\varphi^2\rp, \label{tnads}\\
f(\rho)&=\frac{\rho^2+n^2}{l^2}+\frac{(l^2-4n^2)(n^2-\rho^2)-2m l^2 \rho}{(\rho^2+n^2)l^2}.
\end{align}
It is worth recalling that setting $n=0$, one recovers the hyperbolic AdS black hole \cite{Mann:1996gj,Vanzo:1997gw}. The coordinates $(\theta, \phi)$ parametrize a hyperboloid, and upon appropriate quotients, Riemann surfaces of any genus higher than $1$. For $m\geq-\frac l{3\sqrt3}$, the spacetime contains a black hole, with an event horizon in $\rho_+$, the largest root of the function $f(\rho)$, and a curvature singularity for $\rho=0$. This singularity is timelike when $m<0$, because of the presence of an inner Cauchy horizon, and spacelike otherwise. Saturating the lower bound for the mass parameter one obtains the extremal ground state, with vanishing temperature. On the other hand, for $m=0$ the spacetime is just AdS in a hyperbolic slicing; the horizon is still present and has the interpretation of an acceleration horizon \cite{Vanzo:1997gw}. Upon compactification of the hyperboloid, it becomes a genuine event horizon though.

What happens when we switch the parameter $n$ on? First, it regularizes the metric, in the sense that all its curvature invariants now stay bounded, and the radial coordinate $\rho$ ranges over the whole real axis. But then again, it makes the spacetime prone to CTCs.
The Euclidean section has been studied in \cite{Chamblin:1998pz}, where it was shown that it has a bolt where $f(\rho)$ has its largest root, and no NUT. The $U(1)$ fibration being trivial, there are no Misner strings \cite{Chamblin:1998pz}. Likewise, the Lorentzian spacetime -- on which we focus our interest here -- has no Misner string singularity, but, like for its spherical cousin, there are `large' naked CTCs when $|n|> l/2$ \cite{Astefanesei:2004kn}. The peculiarity of this metric is however that when $|n|\leq l/2$ an event horizon will clothe the CTCs, at least for large enough $m$ \cite{Astefanesei:2004kn}. 
Therefore, for this range of the parameter $n$, causality is preserved in the exterior region of the spacetime, 
and the metric describes a black hole, with an event horizon and an inner Cauchy horizon. The latter encloses a static region of spacetime containing a causality-violating core with CTCs that, contrary to the `large' ones found when $n> l/2$, are contractible and harmless, since they live beyond the Cauchy horizon.

More precisely, let us consider an asymptotic observer in the $\rho>0$ region\footnote{This covers all cases because the Taub-NUT-AdS metric is left invariant under the $\rho\mapsto-\rho$, $m\mapsto-m$ operation.}, and let us define
\BE
m_\pm=\pm\frac l{3\sqrt3}\left|1-12\frac{n^2}{l^2}\right|\sqrt{1-3\frac{n^2}{l^2}}.
\EE
Then, we see that when $0<|n|\leq l/(2\sqrt3)$ the solution is a black hole with two horizons and an innermost CTC core for $m\geq m_-$. When the equality sign holds, the two largest roots of $f(\rho)$ merge and the horizon is extremal. On the other hand,  if the inequality is violated, there are naked CTCs outside the event horizon, for large enough $\theta$. They are confined close to the event horizon, and their radial position $\rho$ cannot be arbitrarily large. For $l/(2\sqrt3)\leq|n|\leq l/2$ we have again a well-behaved black hole if $m\geq m_+$, with two horizons hiding the CTC core, an extremal black hole if $m=m_+$, and naked CTCs if $m<m_-$, with the only difference being that in the range $m_-<m<m_+$ the function $f(\rho)$ as no roots at all\footnote{Note that from the perspective of an observer in the $\rho<0$ asymptotic region, the situation is switched when $m>m_+$ or $m<m_-$: in that region there will be a black hole when there are naked CTCs in the $\rho>0$ region, and naked CTCs when there is a black hole in the $\rho>0$ region. On the other hand, if the mass parameter is such that $m_-<m<m_+$, there are no horizon at all for $l/(2\sqrt3)<|n|\leq l/2$, whereas the CTC core is sandwiched between a $\rho<0$ black hole and a $\rho>0$ black hole when $0<|n|<l/(2\sqrt3)$.}. All these horizons are located at the zeroes of $f(\rho)$ and are Killing horizons.

So, what is the interpretation of the parameter $n$? It is not a NUT parameter, since the spacetime carries no NUT charge. Its effect is to make the metric non-stationary, with $g_{t\phi}\propto n$. As such, observers experience frame dragging, and there is rotation in the spacetime. This is confirmed by the holographic stress tensor \cite{Balasubramanian:1999re}, that can easily be extracted from the metric by changing to Fefferman-Graham coordinates (see e.g.\ \cite{Papadimitriou:2005ii}). The result, not surprisingly, assumes the perfect fluid form,
\BE
T_{ab}=\frac{m}{8\pi G l^2}\left(3u_a u_b+h_{ab}\right),
\EE 
with energy density $\varepsilon=m/(4\pi Gl^2)$ and pressure $\mc P=m/(4\pi Gl^2)$, with the expected equation of state for AdS$_4$ black holes \cite{Awad:1999xx}. Here $u=\p_t$ is a unit timelike vector, and the boundary metric $h_{ab}$ itself is non static \cite{Caldarelli:2012cm},
\BE
\d s^2_{bdy}=-\lp\d t+4n\sinh\lp\frac\theta2\rp^2\d\phi\rp^2+l^2\lp\d\theta^2+\sinh^2\theta\,\d\phi^2\rp,
\label{bdymetric}\EE	
but contains no CTC if $|n|\leq l/2$.
As a result, in addition to the energy density $\varepsilon$, there is a non-vanishing angular momentum density $T_{t\phi}\propto mn$ \cite{Astefanesei:2004kn}. In the coordinates we are using, the angular velocity of the event horizon vanishes, $\Omega_h=0$.  However, this just reflects the coordinate system we are using; in asymptotically AdS spacetimes the conjugate variable to the angular momentum that enters the thermodynamics is the angular velocity of the horizon relative to the boundary $\Omega=\Omega_h-\Omega_{\infty}$ \cite{Caldarelli:1999xj}. The boundary metric \eqref{bdymetric} is indeed rotating, and one finds
\BE
\Omega=-\frac{2n}{l^2+4n^2+(l^2-4n^2)\cosh\theta}.
\EE
We are therefore in presence of a rotating black hyperboloid membrane, as first suggested in \cite{Caldarelli:2012cm}. Since the event horizon has infinite extension, it is not possible to integrate the mass and angular momentum densities to obtain finite charges. Moreover, due to the rotation, it is not possible to compactify the hyperboloid to a smooth Riemann surface to cure the divergence. Also, the black membrane is not rotating uniformly: the rotation is concentrated in the central region of the brane, $\theta\approx0$, with an exponentially vanishing tail. This property explains why these spacetimes can avoid CTCs outside the horizon, and is as well reflected in the mass and angular momentum that are also non-uniformly distributed. Hence, even locally, it is difficult to formulate the first law of thermodynamics, and to check that the entropy follows the Bekenstein-Hawking formula. It is however easy to verify that in the limit of large black holes, $\rho_+/l\gg1$, one recovers the expected hydrodynamic behavior \cite{Bhattacharyya:2007vs}, and with it a local thermodynamic interpretation. Indeed, the Euler relation $\varepsilon+\mc P=T s$ is recovered at leading order in an $l/\rho_+$ expansions, with deviations appearing at the same order as for a Kerr-Newman-AdS black hole (see e.g.\ \cite{Caldarelli:2008ze}), if the entropy density is taken to be $a_h/4G$, with $a_h$ the area density of the horizon.

Further evidence that this is a rotating black membrane comes from a closer look at the Kerr-AdS$_4$ black hole. Surprisingly, this black hole has an ultraspinning limit, typically associated to higher dimensional black holes. Indeed, when taking the rotation parameter $a\rightarrow l$, and simultaneously zooming into the pole in an appropriate way, one reaches a finite limiting metric that is simply \eqref{tnads} with parameter $n=l/2$ \cite{Caldarelli:2012cm}. In higher than four dimensions, the analogous ultraspinning limit of rotating AdS black holes yield rotating black hyperboloid membranes \cite{Caldarelli:2008pz}, and likewise, in four dimensions, the metric \eqref{tnads} describes a rotating black hyperboloid membrane. Moreover, for this particular case (with $n=l/2$) the hyperboloid rotates uniformly $\Omega=1/2l$,  and the boundary always rotates at subluminal speed, $\Omega<1/l$.
Finally, notice that the $n=l/2$, $m=0$ metric is simply AdS in rotating coordinates. Switching on the mass parameter $m$, the metric develops an horizon and becomes a rotating black hole, similarly to what happens with Kerr(-AdS) black holes. On the other hand, keeping $m=0$ and increasing $n$ the geometry deviates from the AdS spacetime contrary to what happens with spherical Kerr(-AdS) metric. This is nevertheless explained by the observation that the $m=0$, $n=0$ metric contains an accelerated horizon.

Summarizing, the parameter $n$ is a \textit{rotation parameter}, analogous to the parameter $a$ appearing in the Kerr metric. At $n=0$, we have a static black hole spacetime with a curvature singularity at $\rho=0$. When $0<n\leq l/2$, the black membrane is put in rotation, and the central spacetime singularity disappears and is replaced by a central CTC core. Finally, when $n>l/2$ the spacetime behaves as a G\"odel spacetime, with large CTCs plaguing it. In some sense, the rotating black membrane is similar to the \textit{under-rotating} BMPV black holes, while the metrics with $l/(2\sqrt3) \leq |n| \leq l/2$ and $m_-<m<m_+$ resemble the \textit{over-rotating} BMPV black holes \cite{Gibbons:1999uv}. It would be interesting to check if these over-rotating menbrane solution also exhibits the `repulson'-like behaviour characteristic of over-rotating black holes \cite{Gibbons:1999uv,Caldarelli:2001iq}. A similar situation is in fact present for the spinning BTZ black hole, where CTCs appear for negative radial coordinate \cite{Banados:1992gq}. The proposed strategy is to cut off the spacetime when one meets the velocity of light surface, at $g_{\varphi\varphi}=0$ (see however \cite{Cornalba:2005je}). Anyway, the inner Cauchy horizon is likely unstable to perturbations, that would replace it with a genuine singularity, cutting off the CTC core. Finally, all these properties survive when a Maxwell field is turned on.

In conclusion, we believe that it would be interesting to investigate further the properties of these black holes, clarify their role in the AdS/CFT correspondence \cite{Caldarelli:2012cm,Mukhopadhyay:2013gja}, and understand their relation to the other family of known rotating hyperbolic black membranes found in \cite{Klemm:1997ea}, and their Taub-NUT generalization \cite{AlonsoAlberca:2000cs}.


\medskip

\textbf{Note added:} During the final writing stage of this project, ref.\ \cite{Bhattacharya:2013hvm} appeared in the {\ttfamily arXiv}, where the authors also report the Taub-NUT-BBMB solution \eqref{NUT-BBMB-g}-\eqref{NUT-BBMB-phi}.

\section*{Acknowledgements}
It is a pleasure to thank Joan Camps, Roberto Emparan, Dietmar Klemm, Harvey Reall, Simon Ross and Robin Zegers for useful discussions. YB would like to thank the SISSA for hospitality during the later stages of this work. This work was partially supported by the ANR grant STR-COSMO, ANR-09-BLAN-0157.
MMC acknowledges support from a grant of the John Templeton Foundation. The opinions expressed in this publication are those of the authors and do not necessarily reflect the views of the John Templeton Foundation.
\appendix
\section{Note on proof of Frobenius condition\label{appendix}}
Consider stationary and axisymmetric metrics of \eqref{action} for $\La=0$ and $\al=0$. 
This means that we assume the existence of two Killing vectors: a Killing vector field $k$ which is asymptotically timelike and a spacelike Killing vector field $m$ whose orbits are closed curves. In addition, we require that they commute, $\left[k,m\right]=0$. It is natural to impose the same symmetries to the scalar field $\phi$, that is $\mc L_k \phi = 0$ and $\mc L_m \phi = 0$ where $\mc L_X$ denotes the Lie derivative with respect to a vector field $X$. In order to write the metric in the Lewis-Papapetrou form, see for example \cite{Straumann:2004fk}, we have to verify that the Frobenius conditions are still true for the above gravitational action (\ref{action}). Indeed, in order to show the Lewis-Papapetrou form in vacuum, or in the presence of a cosmological constant, one has to use the fact that spacetime is an Einstein metric. This is not true here. 
We have to therefore demonstrate that the corresponding one-forms satisfy the Frobenius conditions
\BE k\w m \w \d k=0 \qquad\text{and}\qquad  k\w m\w  \d m=0  \label{frobenius}  \EE
even in the presence of the conformally coupled scalar field $\phi$ in order to simplify the form of the metric\footnote{Thus we correct the argument about the Frobenius conditions given in  \cite{Bardoux:2012nx} where the stationary and axisymmetric problem is studied.}.
Let us consider the twist 1-forms
\BE \om_{(k)} = \frac12 * \lp k \w \d k \rp \qquad\text{and}\qquad  \om_{(m)} = \frac12 * \lp m \w \d m \rp \label{twist}\EE
associated to $k$ and $m$ respectively, with the sign $*$ denoting the Hodge star operator. Therefore the condition  \eqref{frobenius} is equivalent to $i_m \om_{(k)} =0$ and $i_k \om_{(m)} =0$ where $i$ is the interior product. Let us focus on demonstrating the former relation, $i_m \om_{(k)} =0$. 

Since the metric and the scalar field are invariant under the flow of $k$, the equations of motion \eqref{EOMg}-\eqref{EOMphi} infer the following relation after some algebra,
\BE \lp 1- \frac{4\pi G}{3} \phi^2 \rp k_{[a} R_{b]}^{(k)} = 4\pi G \lp k_{[a} \n_b k_{c]} \rp \n^c \phi, \label{Ricci-form}\EE
where we have introduced the Ricci 1-form $R^{(k)}$ with component $R_{ab}k^b$. Then, using component language, the definition \eqref{twist} of the twist form associated to $k$  gives $\ep_{abcd}\om_{(k)}^d =- 3  k_{[a} \n_b k_{c]} $. Moreover, $k$ verifies the identity \cite{Straumann:2004fk}
\BE \d \om_{(k)} = * \lp k \w R^{(k)} \rp, \EE
which, in component language, reads $k_{[a} R_{b]}^{(k)} = - \frac12 \ep_{abcd} \n^c \om^d $. Then, eq.~\eqref{Ricci-form} gives the differential of the twist form associated to $k$,
\BE \d \om_{(k)} =\om_{(k)} \w \d \lp \ln\left| 1-\frac{4\pi G}{3}\phi^2\right| \rp \label{domega} \EE
when $\phi^2 \neq \frac{3}{4\pi G}$. After that, using the Cartan identity $\mc L_m=\d\circ i_m + i_m \circ \d$ and the fact that $\mc L_m \om_{(k)}=0$ (see \cite{Wald:1984rg} for a justification), we have
\BE
 \d i_m \om_{(k)} = - i_m \d \om_{(k)} =
- \lp i_m \om_{(k)}\rp \d \lp \ln\left| 1-\frac{4\pi G}{3}\phi^2\right| \rp +
i_m \lb \d \lp \ln\left| 1-\frac{4\pi G}{3}\phi^2\right| \rp\rb 
\om_{(k)}.
\EE
The second term vanishes in virtue of $\mc L_m\phi=0$,  and we thus obtain
\BE \d\lb \lp 1-\frac{4\pi G}{3}\phi^2\rp i_m \om_{(k)} \rb =0 .\EE
Consequently, $\lp 1-\frac{4\pi G}{3}\phi^2\rp i_m \om_{(k)}$ is a constant function, and the same holds for $\lp 1-\frac{4\pi G}{3}\phi^2\rp i_k \om_{(m)}$. Without loss of generality, we take these constants to be zero since there is a rotation axis on which $m$ has to vanish. Therefore, we have shown that the Frobenius conditions \eqref{frobenius} are true as long as $\phi^2 \neq \frac{3}{4\pi G}$. The situation where $\phi^2=\frac3{4\pi G}$ corresponds to an infinite effective gravitational constant $\tilde{G}$ according to \eqref{Gtilde}.

\bibliography{biblio}
\bibliographystyle{utphys}
\end{document}